\documentclass[12pt]{article}
\usepackage{graphics}

\newcommand{\binom}[2]{\left( \begin{array}{ll} {#1}  \\
                                               {#2}  \end{array}
\right)}
\begin{document}

\title{Market Ecology of Active and Passive Investors}
\author{Andrea Capocci and Yi-Cheng Zhang}
\maketitle
\abstract{We study the role of active and passive investors in an 
investment market with uncertainties. 
Active investors concentrate on a single or a few stocks 
with a given probability of determining the quality of them. 
Passive investors spread their investment uniformly, resembling
  buying the market index. 
In this toy market stocks are introduced as good 
and bad. 
If a stock receives sufficient investment it will survive, otherwise die. 
Active players exert a selective pressure since they can determine to 
an extent the investment quality. 
We show that the active players provide the driving force
whereas the passive ones act as free riders. 
While their gains do not differ too much, we show that the active
players enjoy an edge. 
Their presence also provides better gains to the passive players and 
stocks themselves.}

\section{Introduction}
In the standard finance literature, it is generally stated that
competitive 
financial markets are efficient and an arbitrage opportunity would
instantly 
disappear once smart investors act upon it. This goes under the name
of
Efficient Market Hypothesis (EMH) \cite{Fam} \cite{Fam2}. The proponents never specified
how much a 
probabilistic edge would be reduced and how many smart investors
there 
should be to bring the market to efficiency. The general conclusion
they 
draw from such doctrine is that active investors just waste their
time, 
since nobody can earn
above-market gains. Practitioners, on the other hand, do not heed to
such 
advice and it seems to be impossible to reconcile the academics and 
practice.

Recently an alternative theory is put forward \cite{Zha} \cite{CZ} which maintains
that 
probabilistic edge can never really disappear, but can only be reduced.
There is 
a quantitative relationship  between the reduction and the amount
of investment the active players commit. This is called Marginally
Efficient 
Market (MEM) theory. Under this
general framework in this paper we study the interplay of various
players in 
a toy stock market. Active investors hold relative focus;
passive ones 
just buy everything in the market; as well as the stocks which
can survive or die. In this market ecology we aim to understand the 
interaction among various players, given
the market uncertainties. The salient features coming out from our 
investigation are: the active players provide
the driving force such that the market index can be sustained to
a higher level than random chance would warrant, the passive players enjoy a
free ride. 
In general we find that under quite general conditions active
players 
enjoy an edge over the market index on which the passive players
ride free, 
this provides incentives
to the active players to stay active. On the investment side the
stocks are 
not guaranteed of survival. The more
the active players, the more of the good stocks survive. So in this
ecology 
all players benefit from there being
active players.

\section{ The model}

In a market, investors can choose between two main strategies: either
they focus on some stock, study it in depth and then decide whether to 
invest on it or not, or they can choose to be passive, and invest on all the available stocks
to safely enjoy the average gain (if any) thanks to the central limit
theorem. 
Stocks, on the other hand, can provide positive or negative return, 
and we assume that their survival depends on their capability to 
attract new investments.

We assume that there are $I$ investors and $S$ stocks in which 
they can invest. 
Time is described by the integer variable $t$.
Each stocks $i$ is associated with a spin variable $\sigma_{i}=\pm 1$, that describes 
its quality: if $\sigma_{i} = +1$, the stock $i$ is positive and it is 
good to invest on it, while if $\sigma_{i}=-1$ the stock $i$ is negative and 
one should not invest on it.
An investor can be active or passive.
Each agent can invest at most a unit at each time step.
At each time step, he chooses the stocks to be included in 
his portfolio according to his active or passive strategy, 
then invests an equal amount in each of them. 
We define the {\it gain} $g^{\alpha}$ of a given investor $\alpha$ by
\begin{equation}
g^{\alpha} = \frac{\sum_{i=1}^{S}f_{i}^{\alpha}\sigma_{i}}{P},
\end{equation}
where $f_{i}=1$ if stock $i$ has been included in the portfolio of investor $\alpha$ and
$0$ otherwise, and $P=\sum_{i}^{S}f_{i}^{\alpha}$ is the number of stocks 
included in the portfolio.
In other words, a portfolio gives a positive gain if it includes more good stocks 
than bad ones. 

Let us now define the two strategies.
Passive investors want to take advantage of the average return of the
market. Therefore, they invest a capital $1/S$ on each stock ($f_{i}^{pass}=1$ for all 
$i=1,2,\ldots$) and $P=S$.
On the other hand, each active investor examines a single randomly 
drawn stock and judges it.
Once he has examined a given stock, he invests all his unit on it 
if he thinks it is a good stock, otherwise he does not invest at all.
As we did in \cite{CZ}, we want to represent active investors with 
bounded rationality, who may have a wrong perception of stocks.
We then assume that an active investor who examines a given stock 
has a wrong perception of it with the probability $D$.

When all the investors have made their choice, stocks are selected by 
the amount of the investment they manage to receive. 
Stocks that fail to attract sufficient investment go bankrupt.
Then, a stocks that receives less investment than a fixed threshold 
$T$ is replaced by a new stock. 
The new stock is good with the probability $p_{0}$ and bad with probability $1-p_{0}$.
This process is repeated many times, starting from a random initial condition. 
We denote by $p(t)$ the fraction of good stocks at time $t$.
The {\em market performance} is measured by the average quality of the
market, or {\em index}, $2p(t)-1$.
The system, after a many time steps, falls in a stationary state
where the fraction of good stocks fluctuate around $p^{*}$.

We assume that the fraction of active investors is $\rho$.
Let us suppose, for the time being, that $\rho$ is fixed.
We can compute the average gains of active and passive investors.
An active investor gets a gain of $+1$ if he examines a good
stock and if his perception of the stock is right.
The probability to examine a good stock drawn at random is $p(t)$, 
and the probability to have a right perception of it is $1-D$, then
the gain is equal to $1$ with the probability $p(t)(1-D)$.
The same investor has a negative return, that is, he loses $1$, if he examines a
bad stock, but has a wrong perception of it and decides nevertheless to
invest. Reasoning as above, one can see that this happens with 
probability $(1-p(t))D$.
Otherwise, he receives no return, and this case does not give contribution to the average.
Thus, the average gain of the active investors at time $t$ is $g^{act}=p(t)-D$.
On the other hand, the passive investors enjoy the average behavior of the
whole market and their average gain at time $t$ is $g^{pass}=2p(t)-1$,
If $p(t)<1-D$, which implies $g^{act}>g^{pass}$, the active investors 
enjoy a greater gain than the passive investors.

\section{ The stationary state}

Let $q^{(+)}(t)$ ($q^{(-)}(t)$) be the probability to go bankrupt 
for a good (bad) stock at time $t$.
In the following, we will drop the explicit time dependence (if not necessary) 
to avoid a too heavy notation.
At the stationary state, where the number of good stock replaced by
bad ones equals the number of bad stocks replaced by good ones, we find
\begin{equation}
p^{*}q^{(+)}(1-p_{0}) = (1-p^{*})q^{(-)}p_{0};
\end{equation}
From this, we obtain
\begin{equation} \label{stat}
p^{*} = p_{0}q^{(-)}  / [(1-p_{0})q^{(+)} + p_{0} q^{(-)}].
\end{equation}
Let us consider a given bad stock and check its total 
investment received to determine its fate.
$q^{(-)}$ is the probability that a given bad stock receives less 
than $T$ units of investment.
All stocks, bad or good, enjoy the same investment coming from 
the passive investors, spread out over all the market.
This amount per stock is a constant equal to $T_0 = I(1-\rho)/S$.

\begin{figure}[t]
\scalebox{0.8}[0.8]{\rotatebox{270}{\includegraphics{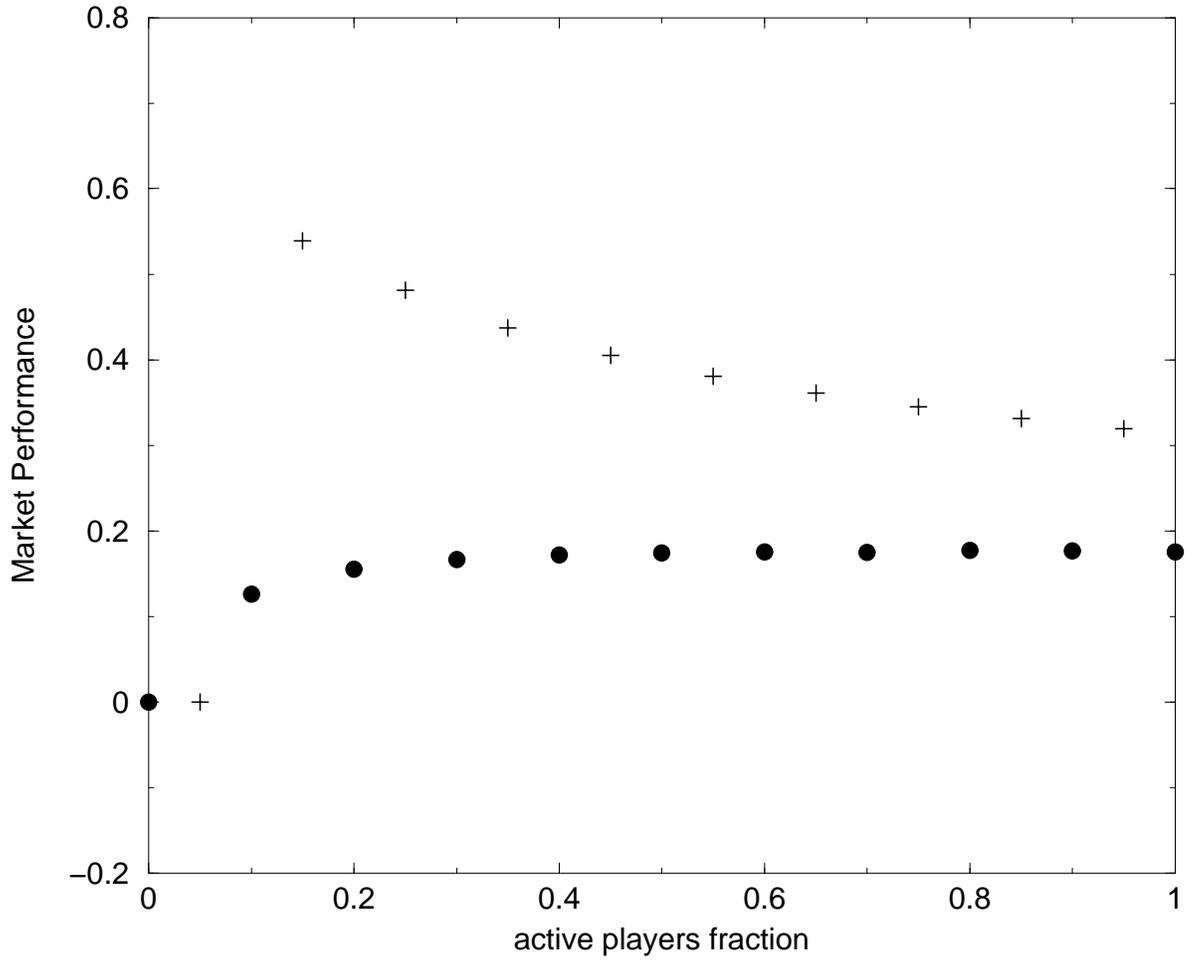}}}
\caption{Market Performance at the stationary state as a function of the active investor fraction in a model with $I=1000$, 
$S=100$, $p_{0}=0.5$, $T = 11$ (circles) and $T = 9$ (plus), according to equation \ref{edge}}
\label{fig1}
\end{figure}

To fill the gap between $T_{0}$ and the threshold $T$, the bad 
stock should receive sufficient investment from the active 
investors who mistake the bad stock for good.
The probability that $k$ active investors invest on a bad
stock obeys a binomial distribution.
Each active investor examines a stock with probability $1/S$.
If it is a bad stock, the investor invests on it if he mistakes 
it, which happens with the probability $D$. 
Then, the probability of receiving a unit of investment from each 
active investor is $D/S$.
The probability to receive $k$ units of investment from all active investors
is $b(k,\rho I,D/S)$, where $b(n,M,z)$ is the usual binomial 
probability of observing $n$ events on $M$ trials, 
if the probability of the single event is equal to $z$, i.e.
\begin{equation}
b(n,M,z) \equiv \binom{M}{n}z^{n}(1-z)^{M-n}.
\end{equation}
Then, the probability for a bad stock to go bankrupt is
\begin{equation} \label{minus}
q^{(-)} = \sum_{k=0}^{T-T_{0}}b(k,\rho I, D/S).
\end{equation}
Likewise, the computation of the bankrupt probability for a good
stock $q^{(+)}$ gives
\begin{equation} \label{plus}
q^{(+)} = \sum_{k=0}^{T-T_0}b(k,\rho I,(1-D)/S).
\end{equation}
By means of the binomial distribution computed above, we
can evaluate the average investment received by each stock.
From active agents, each good stock receives on average $(1-D) \rho I/S$ 
and each bad stock $D\rho I/S$.
By adding $T_{0}$ to both quantities, we obtain that each good stock
receives on average $(1-D\rho)I/S$ units of investment
and each bad stock receives $[1-D(1-\rho)]I/S$.
By replacing eqs. (\ref{minus}) and (\ref{plus}) in equation (\ref{stat}), 
we obtain
\begin{equation}\label{p_rho}
p^{*} = p^{*}(\rho) = \frac{p_{0}\sum_{k=0}^{T-T_0}b(k,\rho 
I,D/S)}{\sum_{k=0}^{T-T_0}[(1-p_{0})b(k,\rho 
I,D/S)+p_{0}b(k,\rho I,(1-D)/S)]}.
\end{equation}
The above formula can be easily verified by simulations.
For different values of $T$ we obtain qualitatively
different behaviors.
If $T>I/S$, i.e. if the market is under severe selective pressure, 
the market index increases monotonically as a function of the number 
of active investors.
For low values of $\rho$, i.e. when active investors are rare, all
stocks receive about the same amount of investment, since the passive 
agents invest the same quantity $1/S$ on each stock regardless its return.
Thus, for $\rho = 0$, each stock receives $I/S<T$ investment and no one survives:
\begin{eqnarray}
q^{(+)} & = & 1,\\
q^{(-)} & = & 1.
\end{eqnarray}
and 
\begin{equation}
p^{*}=p_{0}.
\end{equation}
As $\rho$ (the fraction of active investors) increases, investment is
allocated in a more selective way, since the active investors can distinguish
to a certain extent good from bad stocks.
Thus, if there are more active investors, good stocks have a higher 
probability to receive the minimal $T$ investment and then survive.
The presence of the active investors then gives an edge to the good
stocks, and $p$ increased with $\rho$ as shown in figure \ref{fig1}.
When $D \rightarrow 0$ and $T=I/S$, the active investor exercise the strongest
selective pressure on the market (provided that $T>I/S$), since 
they invest with great accuracy and the threshold is exactly the 
average investment received by all stocks.
In this situation, the number of good stocks that receive less than
$T$ investment is the minimal.
We now ask whether the selective pressure on the market can so strong
that passive investor enjoy a greater gain than the active ones.
This happens if $p^{*}>1-R$ in a market where active agents invest with a 
very high precision, that is, $D \rightarrow 0$.

Since $p^{*}$ is an increasing function of $\rho$, we check if
the condition is verified for $\rho=1$.
Replacing $\rho=1$, $D \simeq 0$ and $T=I/S$ in equation (\ref{p_rho}), we obtain
\begin{equation} \label{edge}
p^{*} \simeq \frac{p_{0}\sum_{k=0}^{I/S}\delta(k)}{(1-p_{0})\sum_{k=0}^{I/S}\delta(k) + p_{0}\sum_{k=0}^{I/S}b(k,I,1/S)}.
\end{equation}
The $\delta$ functions in equation (\ref{edge}) come from the limit
\begin{equation}
\lim_{z \rightarrow 0} b(n,M,z) = \delta(n).
\end{equation}
Since 
\begin{equation}
\sum_{k=0}^{I/S}\delta(k)=1,
\end{equation}
and
\begin{equation}
\sum_{k=0}^{I/S}b(k,I,1/S)\simeq \frac{1}{2},
\end{equation}
we have, from equation (\ref{edge}),
\begin{equation}
p^{*}(\rho=1) \simeq \frac{2p_{0}}{2-p_{0}}.
\end{equation}
Thus, the condition $p^{*} > 1-D$ can be satisfied only if $p_{0}>2/3 + \mathcal{O}(\epsilon)$.
If $T<I/S$, we observe a radically different behavior.
If there is no active investors ($\rho=0$), each stock receives 
$I/S>T$ investments, thus all stocks survive and we have
\begin{equation}
q^{(+)}=q^{(-)}=0.
\end{equation}
This remains true until the fraction of active investors is such that
\begin{equation} \label{rho_c}
(1-\rho)I/S \geq T.
\end{equation}
Then, there is a critical fraction of investors $\rho_{c}$ such that if
$\rho < \rho_{c}$ all stock receive more than the threshold investment.
Equation (\ref{rho_c}) implies $\rho_{c} = 1 - ST/I$.
Beyond this value, the investment coming from the passive agents is no 
more sufficient to keep all the investors above the threshold $T$.
As the number of active investors increases over the value $\rho_{c}$,
investment move from bad to good stocks, and the bankrupt probability
for a bad stocks grows much more than for the good stocks, as one can see in 
figure \ref{fig1}.
For higher values of $\rho$, however, the fraction of good stocks $p$ 
starts to decrease.
Indeed, when there are many active players bad stocks can take advantage 
of their more frequent errors and the survival probability for a bad stock 
slightly increases, as it is shown in figure \ref{fig1}.

\section{Effort}
Let us now define the {\em effort} $E$. 
We use the same notation of ref. \cite{CZ} and assume that the error 
probability $D$ is a decreasing function of the effort; we choose 
for simplicity 
\begin{equation} \label{D_Eff}
D = 1/(E+2).
\end{equation}
Thus, $E$ is positive definite and gives a measure of the skill of 
the active investors, which is assumed to be constant for all 
the active investors.
If the active investors provide no effort ($E=0$) they would invest 
at random.
Error would then be the largest ($D=\frac{1}{2}$).
On the other hand, a perfect perception of the market cannot be
reached by a finite effort, since $D=0$ corresponds to 
$E=\infty$.
We want to investigate the role of the effort in a market, where the 
fraction of active investors is fixed and the system is in 
the stationary state.
The active investors' ability to distinguish good stocks from
bad ones is a function of the effort they provide.

\begin{figure}[t]
\scalebox{0.8}[0.8]{\rotatebox{270}{\includegraphics{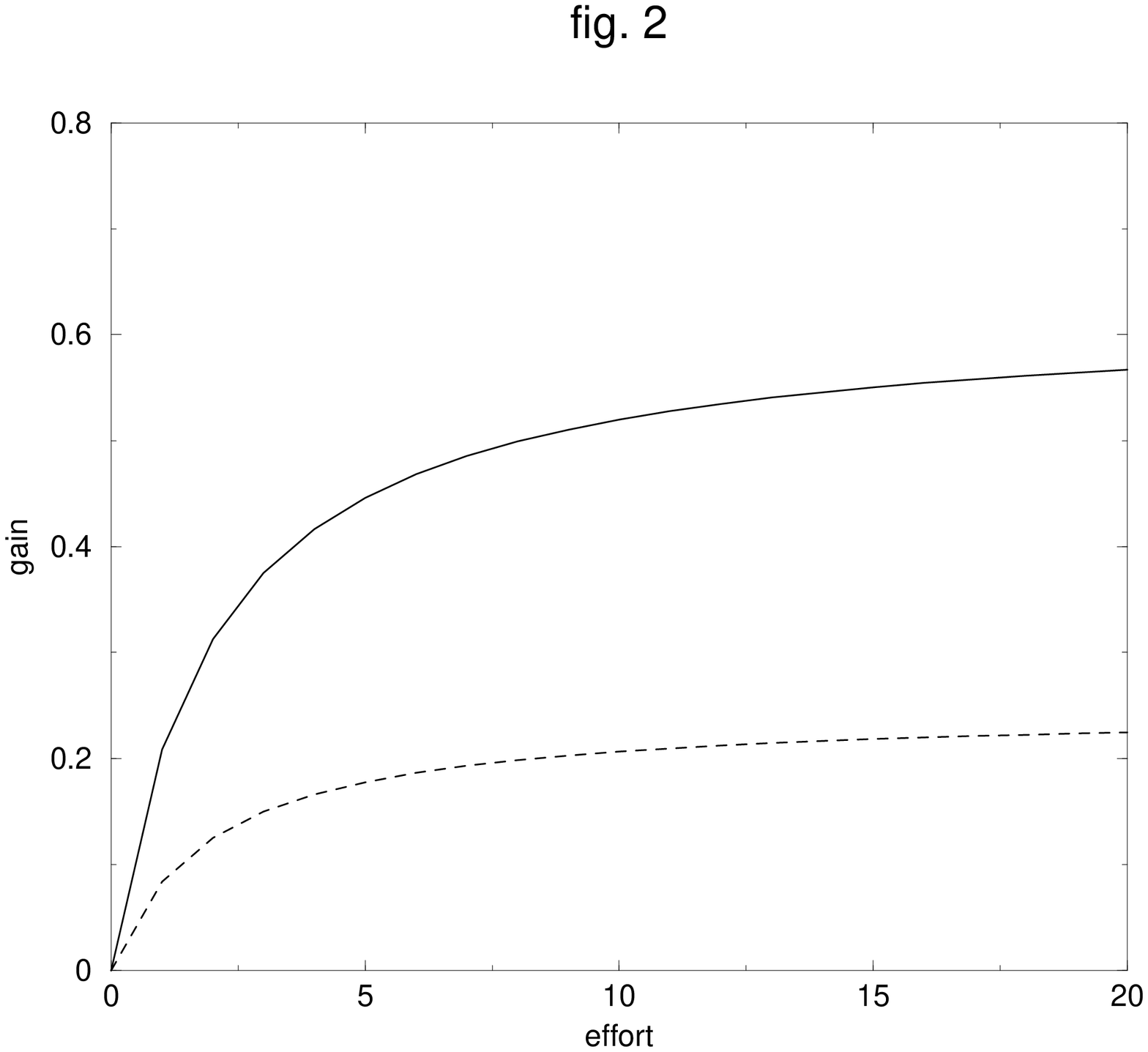}}}
\caption{Average Gain of active (solid line) and passive (dashed line) investors as a function of the 
effort $E$ of the active investors, in a model with $\rho = 0.1$, $I=500$, $S=100$, $T=5$, $p_{0}=0.5$}
\label{fig2}
\end{figure}

Let us consider the gain of an investor as a function 
of his effort with $\rho$ fixed.
By replacing $p(\rho)$ in the expressions of the gains $g^{act}$ and 
$g^{pass}$, one can show that both the gains of the active and passive 
investors grow monotonically with $E$. This behavior is shown in figure \ref{fig2}. 

We conclude that the effort by the active investors exercises 
a selective pressure on the market performance.
We notice that if the active investors provide the minimal
effort $E=2$, which corresponds to an error probability $D=1/2$, the
number of bad and good stocks weeded out of the market are equal, 
since both kinds of investors are indifferent to the return of the stock 
on which they are investing.

We then generalize, by numerical simulation, our study of the market
assuming that the active agents invest on more than only one stock.
As we did in ref. \cite{CZ}, we assume that the missing information,
measured by the error probability $D$, grows with the
number of stocks examined by the active investors, $N$ and that a
residual ignorance $D_{0}$ remains also for $N\rightarrow 0$, i.e. 
\begin{equation}
\label{err_N} D(N)=D_{0}+A\frac{N}{E}, 
\end{equation} where $A$ is a constant determined by imposing $D(S)=1/2$.
This means that an investors who examine all stocks actually behaves like
a random investor.

\begin{figure}[t]
\scalebox{0.8}[0.8]{\rotatebox{270}{\includegraphics{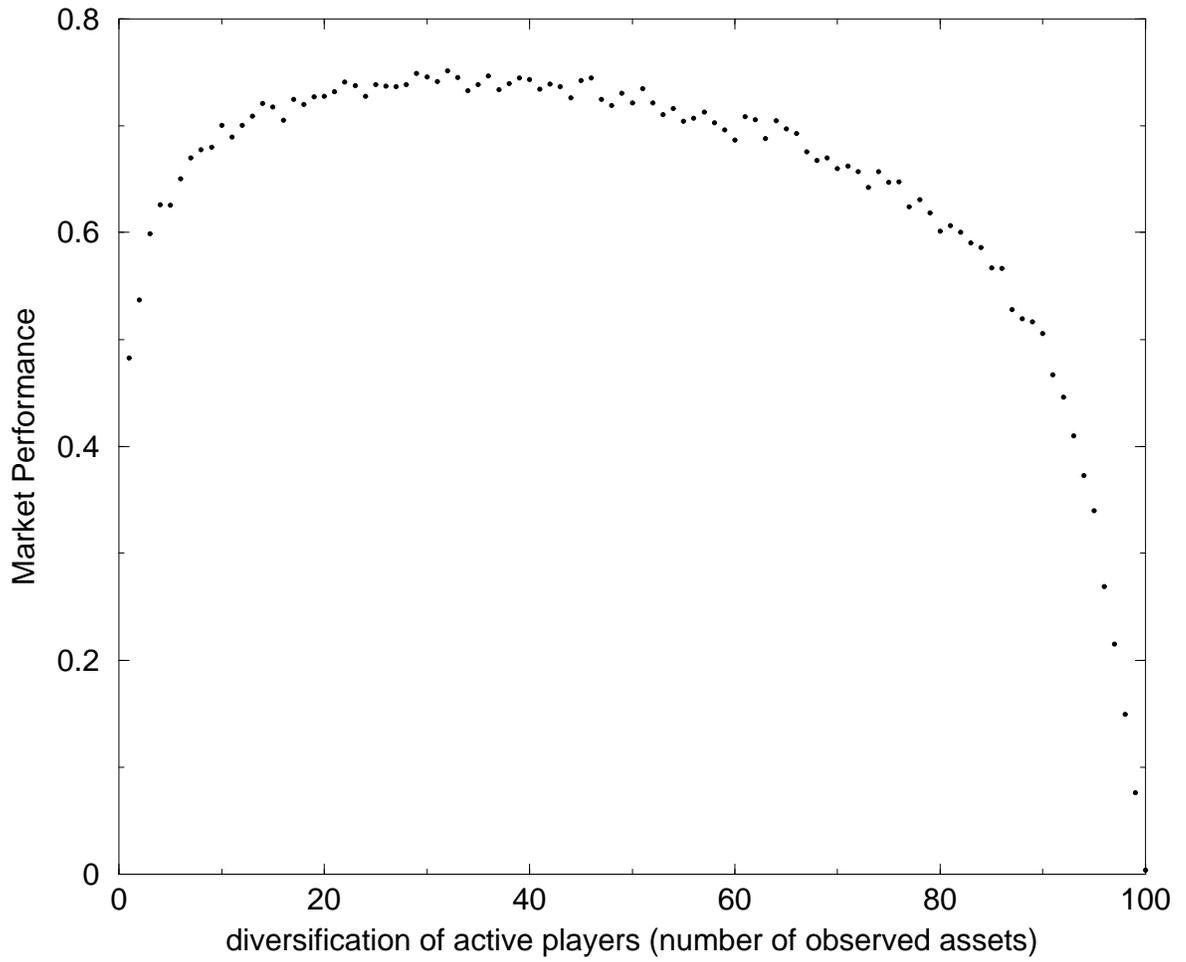}}}
\caption{Market performance as a function of the diversification of active strategies with $\rho = 0.2$, 
$I=500$, $S=100$, $T=5$, $p_{0}=0.5$}
\label{fig3}
\end{figure}

We see that the market performance of an active investor $2p-1$ 
has a maximum for a finite value $0<N^{opt}<S$, as it is shown in figure \ref{fig3}.

Indeed, the of information needed to include more stock in a portfolio
then a tradeoff has to take place between the need of diversification and 
the need of including {\em good} stocks in a portfolio.
This confirms the conclusions of \cite{CZ} and \cite{Zha}, where the diversification 
of a portfolio is found to be bounded by a finite cost of information about 
the stocks, which acts as a source of inefficiency in a market.

\section{Complex behavior of self organized investors}

Let us now suppose that the passive investors do not invest in all stocks,
but they choose at random the stocks to be included in their portfolio.
This is the easiest strategy to follow: such a noise trader chooses its
portfolio at random and invests on each stock with probability $1/2$, regardless
the stock is good or bad.
Once he has chosen the stocks, he invests an equal amount on each of them.
As above, the passive investors have a unit to invest, and they invest their 
entire capital.
An investor who follows this strategy enjoys the average return of the market, 
after many time steps.

On the other hand, an active investor follows the same strategy as in the 
previous model: he examines one single stock drawn at random and, if he thinks
that it is a good stock, he invests on it.

We now allow investors to adapt and change strategy when their last 
investment gives a non positive gain, so that the density of active
investors $\rho(t)$ changes with time: if an investor enjoys a null or negative 
gain, he switches to the other strategy with the probability $\rho_{0}$.
As above, we study the properties of the system in the stationary state, 
where $\rho(t)$ and $p(t)$ start oscillating around their stationary values $\rho^{*}$ 
and $p^{*}$ respectively.

By numerical simulation shown in figure \ref{fig4}, we observe that the
fraction of passive investors $1-\rho^{*}$ at the stationary state 
increases with the effort of the active investors.
Likewise, the market performance increases with the effort of the active investors, 
as do the average gains of active and passive investors.

This can be qualitatively interpreted by a simple argument.
The active investors have a clear advantage if their effort is
increased, since they invest in the market with a higher precision.
But the passive investors as well take advantage of the effort: the better the active
investors work, the easier is the life for the passive investors, so that their
population increases.
If there are many active investors, the market becomes more selective and
the fraction of good stock increases.
This allows more passive strategies to reach a better results and grow in number.
As a non trivial results, it appears that the more the active investors 
are successful, i.e. invest with a greater accuracy, the more the passive
investors are favored.

\section{Approximated analysis of the self organized model}

The properties at the stationary state can be approximately
computed.
We study the stationary state of the system as a function of the
effort provided by the active investors, whose expression, from equation (\ref{D_Eff}) 
can be written as $E=D^{-1}-2$.
We note by $\tau_{a}$ the probability that an active strategy
obtains a negative or null gain, and with $\tau_{f}$ the same probability for
an average strategy.
At the stationary state, the numbers of agents who switch from active to 
passive strategy and from passive to active strategy must be the same.

The probability that an active investor switches to a passive strategy
is equal to $\tau^{act}(p)=D(1-\rho_{0})(1-p)$, i.e. the product of the probability $(1-p)$ of examining 
a bad stock, the probability $D$ of having a wrong perception of that stock and
the probability $1-\rho_{0}$ of replacing an active strategy with a passive one.

On the other hand, a passive investor does not include a fixed number of stocks in his
portfolio.
Let us assume that he invests on $n$ stocks. 
In such a case, the probability to include $m$ bad stocks is a 
binomial one, $b(m,n,p)$.
Thus, for a given $n$, the probability to obtain a negative gain is
$\tau^{pass}(p)=\sum_{m<n/2} b(m,n,p)$.
The probability to invest over $n$ stocks drawn at random obeys a
binomial distribution, $P(n)=b(n,S,1/2)$.
Then, we can average $\tau^{pass}$ over all values of $n$, and we
obtain
\begin{equation}
\tau^{pass}(p) = \sum_{n}P(n)\sum_{m<n/2}b(m,n,p).
\end{equation}
By imposing the detailed balance condition
\begin{equation}
(1-\rho) \tau^{pass}(p) = \rho \tau^{act}(p),
\end{equation}
we can compute the active investors fraction at the stationary 
state, $\rho^{*}$ as a function of the fraction
of good stocks at the stationary state, $p^{*}$.
Indeed, he stationary state condition implies

\begin{equation} \label{rho_p}
\rho^{*} = \frac{\rho_0 \tau^{pass}(p^{*})}{(1-\rho_{0}) \tau^{act}(p^{*}) + \rho_{0} \tau^{pass}(p^{*})}.
\end{equation}
One can proceed as above to write a detailed
balance equation for the stocks.
A given stock is replaced if it receives investment less than $T$.
As seen above, the number of stocks included in its portfolio obeys a binomial distribution with mean value $S/2$.
Let us assume that all the passive investors invest on $S/2$ stocks, investing
$2/S$ on each stock.
Under this assumption, a stock included in the portfolio of $Sx/2$ 
passive investors receives from them an investment equal to $x$.
Thus, the probability $Q_{<}(m)$ that a given stock receives from the passive agents an 
investment less than $m$ is equal to the probability to be included in the portfolio of less than 
$Sm/2$ passive agents over $(1-\rho)I$.
The probability to be included in the portfolio of $k$ passive investors
is $b(k,(1-\rho)I,1/2)$.  
Thus, the probability that a given stock receives less than $m$ investment from the passive 
investors is equal to 
\begin{equation}
Q_{<}(m) = \sum_{k=0}^{Sm/2}b(k,(1-\rho)I,\frac{1}{2}).
\end{equation}
The probability $Q_{<}(m)$ does not depend on the quality of the stock, 
since this investment comes from the agents who invest at random on 
good and bad stocks.
The active investors, conversely, provide an effort $E$ in examining
the quality of stocks.
The probability that a given good stock receives an investment from an active 
agent is $(1-D)/S$, where $1/S$ is the probability to be examined by 
the active investor and $(1-D)$ is the probability to be exactly judged,
i.e. that the investor has a correct perception of the good
stock and invests on it.
Therefore, the probability that a good stock receives $l$ investments from
the $\rho I$ active investors is $R^{(+)}(l) = b(l,\rho I,D/S)$.
Likewise, for a bad stock we obtain $R^{(-)}(l) = b(l,\rho I,(1-D)/S)$.
Then, the probability that a stock receives less than $T$ is
\begin{equation}
q^{(+)} = \sum_{l=0}^{T}R^{(+)}(l)Q_{<}(T-l),
\end{equation}
and
\begin{equation}
q^{(-)} = \sum_{l=0}^{T}R^{(-)}(l)Q_{<}(T-l),
\end{equation}
for good and bad stocks.
We notice that these two probabilities are functions of $\rho$.
They allow us to write
\begin{equation}
p^{*}=p^{*}(\rho^{*})=\frac{p_{0}q^{(-)}(\rho^{*})}{p_{0}q^{(+)}(\rho^{*})+(1-p_{0})q^{(-)}(\rho^{*})}
\end{equation}
and, by replacing $\rho^{*}=\rho^{*}(p^{*})$ from equation (\ref{rho_p}), 
we obtain a self-consistent equation for $p^{*}$, that can be solved 
numerically but exactly.
As a consequence, we can plot $p^{*}$ and $\rho^{*}$ as functions 
of the effort $E$. 
The results are verified by numerical simulation.

\section{Conclusion}
We describe the statistical properties of a simplified market model composed by investors and stocks.
Stocks can only be {\em good} or {\em bad}, according to the sign of their return.
Investors can be {\em active}, i.e. study the stocks and invest on the positive ones,
or {\em passive}, that is, behave as free riders and take advantage of the market performance, without any selective effort.
We investigate the stationary state of this model. 
First we consider a fixed population of active investors, and we select stocks according to the investment they manage to attract.
We observe by numerical and analytical means that increasing the fraction of active investors improve the market performance, which can be exploited as well by passive investors.
Second, we consider a model in which investors can switch from active to passive strategy and {\em vice versa}.
We observe that as the accuracy of active investors is increased, the population of passive investors grows.
We can then conclude that active and passive investors live in a sort of {\em symbiosis}, although the two
 kinds of investors are in competition (selective force {\em vs} diversification).
Indeed, the gain of the active agents, due to an increased accuracy, can result in a greater gain 
for the free riders as well, due to a better performance of the whole market.

\begin{figure}[t]
\scalebox{0.8}[0.8]{\rotatebox{270}{\includegraphics{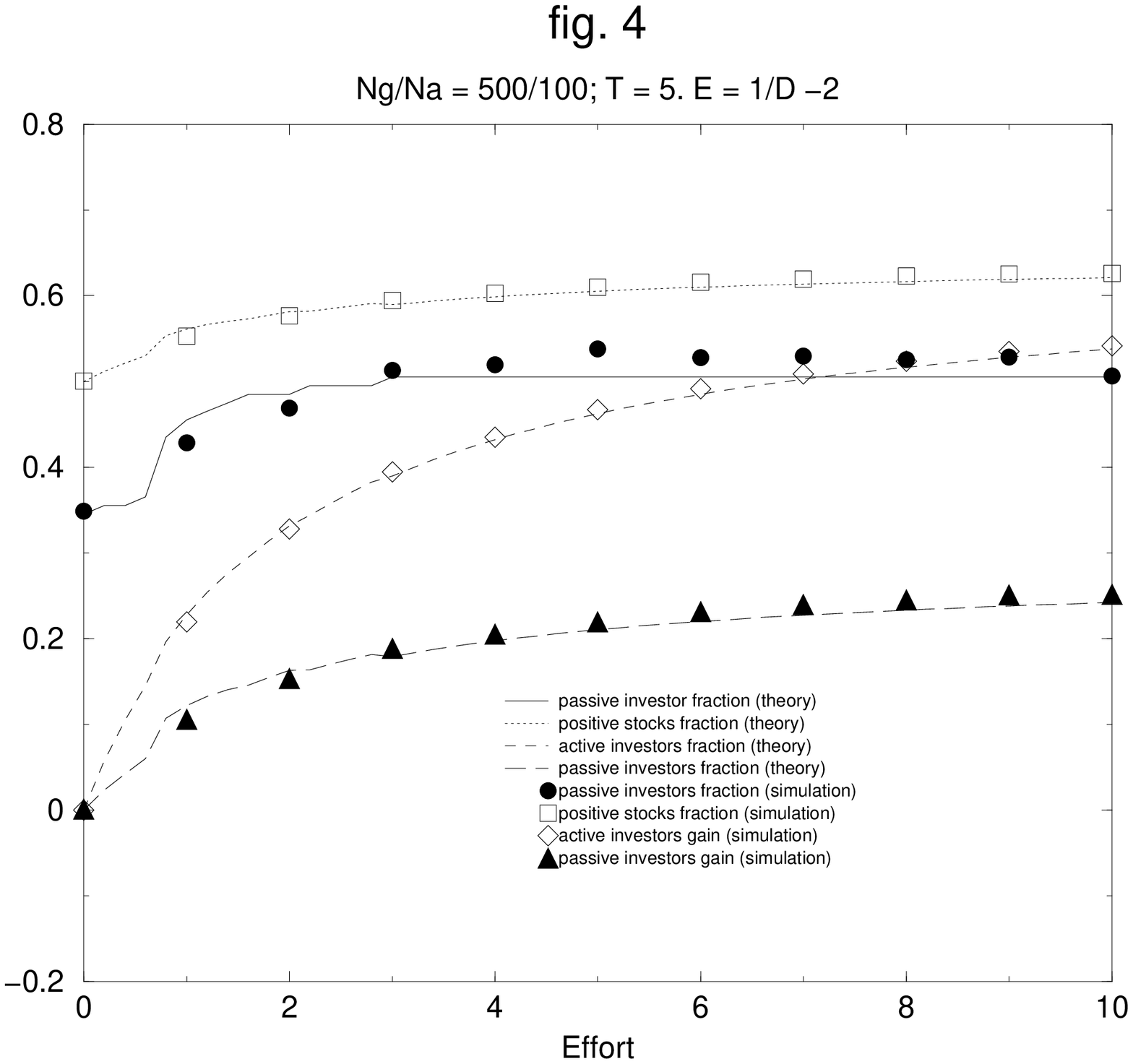}}}
\caption{Passive investors fraction (solid line: theory, circles simulation), 
positive stock fraction (short dashed line,squares), active (dashed line, diamonds) and 
passive (long-dashed line, triangles) investor fraction as a function of the effort $E$ in 
the self organized model, with $I=500$, $S=100$, $T=5$, $p_{0}=0.5$}
\label{Fig4}
\end{figure}

\end{document}